\documentclass[preprintnumbers,amsmath,amssymb]{revtex4}
\usepackage{graphicx}
\usepackage{amsmath}\usepackage{slashed}
\usepackage{color}
\usepackage{mathtools}
\usepackage{txfonts}

\UseRawInputEncoding
\begin{document}

\thispagestyle{empty}
\title{Quantum Theory and Unusual Dielectric Functions of Graphene}

\author{
V.~M.~Mostepanenko}
\affiliation{Central Astronomical Observatory at Pulkovo of the Russian Academy of Sciences, St.Petersburg,
196140, Russia}
\affiliation{Peter the Great Saint Petersburg
Polytechnic University, Saint Petersburg, 195251, Russia}

\author{
G.~L.~Klimchitskaya}
\affiliation{Central Astronomical Observatory at Pulkovo of the Russian Academy of Sciences, St.Petersburg,
196140, Russia}
\affiliation{Peter the Great Saint Petersburg
Polytechnic University, Saint Petersburg, 195251, Russia}

\begin{abstract}We address the spatially nonlocal dielectric functions of graphene
at any frequency derived starting from
the first principles of thermal quantum
field theory using the formalism of the polarization tensor.
After a brief review of this formalism, the longitudinal and transverse
dielectric functions are considered at any relationship
between the frequency and
the wave vector. The analytic properties of their
real and imaginary parts are investigated at low and high frequencies.
Emphasis is given to the double pole at zero frequency which
 arises in the transverse dielectric function. The role of this
unusual property for solving the problem  of disagreement between
experiment and theory in the Casimir effect is discussed.
We { {guess}} that a
more complete dielectric response of ordinary metals should also be spatially nonlocal
and its transverse part { {may}} possess the double
pole in the region of evanescent waves.
\end{abstract}

\maketitle

\newcommand{\ve}{\varepsilon}
\newcommand{\vq}{v_F^2q^2}
\newcommand{\svq}{\sqrt{v_F^2q^2-\omega^2}}
\newcommand{\wsvq}{\sqrt{\omega^2-v_F^2q^2}}
\newcommand{\oq}{(\omega,q)}
\newcommand{\boq}{(\omega,\mbox{\boldmath$q$})}
\newcommand{\dxe}{\frac{dx}{e^{\,\beta x}+1}}

\section{Introduction}

It is common knowledge that quantum physics originated in 1900 from the work of
Max Planck who derived the distribution law for the monochromatic radiation, introduced
a concept of the quantum of energy and the new fundamental constant $h$ now known
as the Planck constant \cite{1I}. In 1905, Albert Einstein arrived at a concept of the quanta
of light, which were later called photons, and explained on this basis the photoelectric
effect \cite{2I}. However, the first quantum theory, quantum mechanics, was created one hundred
years ago by Werner Heisenberg (1925) \cite{3I} and Erwin Sch\"{o}dinger (1926) \cite{4I}.
The relativistic quantum mechanics formulated by Paul Dirac in 1928 \cite{5I} introduced the
concept of antiparticles and opened a way to the formulation of quantum field theory and
its applications to all fundamental interactions of nature during the twentieth century.

For a long time, it was believed that quantum theory is only needed to describe very small
submicroscopic objects on atomic and even subatomic scales. Later it was understood,
however, that there are a lot of macroscopic quantum phenomena characterized by the
scales greatly exceeding the atomic ones. These are the superconductivity, superfluidity,
quantum Hall and Josephson effects, Bose-Einstein condensation, the Casimir effect etc.
Currently the macroscopic quantum phenomena are not of only an academic interest,
but are widely used in many technological and industrial applications, as well as in
metrology.

Considerable recent attention has been focussed on various novel materials
described by quantum theory, and the
two-dimensional sheet of carbon atoms called graphene occupies a prominent place among
them. Graphene is remarkable for many reasons including its unique mechanical, electrical
and optical properties \cite{1,2,3}. At energies below approximately 3~eV, graphene is well
described by the Dirac model. This means that the field of either massless or very light
electronic quasiparticles in graphene satisfies the (2+1)-dimensional Dirac equation rather
than the Schr\"{o}dinger equation which describes the standard quasiparticles considered in
condensed matter physics. In doing so, the Fermi velocity in graphene $v_F \approx c/300$
plays the same role as the speed of light $c$ in the usual Dirac equation.

{The response of graphene to} the electromagnetic field
is spatially nonlocal. It is commonly described by the tensors of electric conductivity
or dielectric permittivity. For a graphene sheet in the absence of constant magnetic field,
these tensors are characterized by the two functions each depending on frequency $\omega$,
the two-dimensional wave vector ${\bf q}$ and on temperature $T$ (for the gapped and doped
graphene they also depend on the mass gap parameter and chemical potential). These are the
longitudinal $\sigma^{\rm L}$ and transverse $\sigma^{\rm T}$ conductivities and the corresponding
dielectric functions $\varepsilon^{\rm L}$ and $\varepsilon^{\rm T}$. In the case of two spatial
dimensions, the dielectric functions are expressed in terms of conductivities as \cite{4,5}
\begin{equation}
\varepsilon^{\rm {L,T}}\boq=1+\frac{2\pi iq}{\omega}\sigma^{\rm {L,T}}\boq,
\label{eq1}
\end{equation}
where $q=|\mbox{\boldmath$q$}|=\sqrt{q_1^2+q_2^2}$.
This means that in the Gaussian units used here and below the conductivities of graphene
have the dimension of cm/s (for the three-dimensional materials,  the dimension of
conductivity is 1/s).

The response functions of graphene were investigated in both the spatially  local
($\mbox{\boldmath$q$}=0,~\varepsilon^{\rm L}=\varepsilon^{\rm T}=\varepsilon$) and nonlocal cases using
a  number of more or less phenomenological approaches such as the hydrodynamic
model \cite{5a,5b,5c}, Boltzmann transport theory and the Drude
model \cite{11,11a,11b,13,14,18,24,25,31}, current-current correlation functions and the
random phase approximation \cite{9,12,12a,18a,21,23a,26,26a,26b,29a}, density-functional
theory \cite{23,23aa}, Kubo response theory \cite{6,7,7a,7ab,8,15,16,17,27,27a,28,29},
modeling graphene optics in terms of Lorentz-type oscillators \cite{31a} and by using
the Fresnel reflection coefficients \cite{32}, etc. (see also the reviews \cite{34,35,36}).
 The obtained results are of different levels
of accuracy and areas of application. In the framework of the Dirac model, however,
the response functions of graphene can be found
 {exactly} starting from the first principles of thermal quantum field theory
using the formalism of the polarization tensor in (2+1)-dimensions.

In this article, we discuss the properties of the longitudinal and transverse dielectric
functions of graphene expressed in the framework of the quantum field theory
via the components of the polarization tensor.
The dependence of these functions on frequency is investigated over the entire region of positive
frequencies (the dependence of the dielectric function of graphene on temperature is
investigated in Ref.~\cite{39a}).
 It is shown
that these functions possess some usual properties characteristic of common materials.
Thus, they satisfy the Kramers-Kronig relations, go to unity with an indefinitely increasing
frequency, and have the positive imaginary parts as it must be in accordance with the
second law of thermodynamics \cite{37}. At the same time, we demonstrate that the
transverse dielectric function of graphene possesses an unusual property by
having the double pole at zero frequency   (it is generally believed that
at zero frequency the response functions of metallic and dielectric materials have a
single pole and are regular, respectively).   We propose that in the region of evanescent waves the
transverse dielectric function of ordinary metals { {may} have the
double pole as well.

We start in Section 2 with a brief review of the most necessary results
 regarding the polarization tensor of graphene.
Then  in Section 3 we consider the pro\-per\-ties of dielectric functions of graphene expressed via the
polarization tensor at  low  frequencies.
Section 4 is devoted to the case of high frequencies. Sections 5 and 6 contain the discussion and
our conclusions.
 For the sake of clarity in presentation,
all mathematical equations are written for the case of a pristine graphene possessing the
zero mass gap parameter and chemical potential. However, all the results
presented below remain valid for the gapped and doped graphene sheets.

\section{Polarization tensor of graphene}

In the one-loop approximation, the interaction of electronic quasiparticles in graphene
with the electromagnetic field is described by the quasiparticle loop diagram having two
photon legs. It is represented by the polarization tensor $\Pi_{\mu\nu}$, where
$\mu,\nu=0,1,2$. At zero temperature, the polarization tensor has long been
calculated within (2+1)-dimensional quantum field theory \cite{38,39}. Specifically
for graphene, whose properties are temperature-dependent, the polarization tensor was
studied in detail at both zero and nonzero temperature \cite{6,40,41,42,43,44}. In
the latter case, the formalism of thermal quantum field theory in the Matsubara
formulation has been used.

The expressions for the polarization tensor of graphene valid over the entire plane of
complex frequency, including the real frequency axis, were obtained in \cite{45,46}
(the previously obtained expressions \cite{44} are valid only at the pure imaginary
Matsubara frequencies). They were used for investigation of the electric
conductivity \cite{47,48,49,50} and reflectivity \cite{51,52,53} of graphene, as well as
of the Casimir and Casimir-Polder forces in out-of-thermal-equilibrium graphene
systems \cite{54,55,56,57,58,59}.

All components of the polarization tensor can be expressed via the two independent
quantities \cite{44}, e.g., via $\Pi_{00}$ and
\begin{equation}
\Pi\boq=q^2\Pi_{\mu}^{\,\mu}\boq+\left(\frac{\omega^2}{c^2}-q^2\right)
\Pi_{00}\boq .
\label{eq2}
\end{equation}
\noindent
There are different but mathematically equivalent representations for the
quantities $\Pi_{00}$ and $\Pi$. Below we use that ones
presented in Ref.~ \cite{54}.

{ {Note that the entire range of positive frequencies from zero
to infinity can be divided into the regions of evanescent, $0<\omega <cq$,
and propagating, $\omega \geqslant cq$, waves. In its turn, in the region of
evanescent waves, it is convenient to separate the subregion of strongly
evanescent waves, $0<\omega <v_Fq$.}}
The explicit expressions for $\Pi_{00}$ and $\Pi$ have different forms in the
regions $0<\omega<v_Fq$ and $\omega>v_Fq$. We start with the region
$0<\omega<v_Fq$, { {i.e., with the strongly}} evanescent waves. In this region,
for $\Pi_{00}$ one has \cite{54}
\begin{eqnarray}
&&
{\rm Re}\,\Pi_{00}\boq=\frac{\pi\alpha\hbar c q^2}{\svq}+
\frac{8\alpha\hbar c}{v_F^2}\left\{
\vphantom{\left[\int\limits_{0}^{v_Fq-\omega}\dxe\right]}
\frac{\ln2}{\beta}\right.
\label{eq3} \\
&&\left.
+\frac{1}{2\svq}\left[\int\limits_{0}^{v_Fq-\omega}\!\!\!\!\dxe F_1(x)
-\!\!\!\int\limits_{0}^{v_Fq+\omega}\!\!\!\!\dxe F_2(x)\right]\right\}
\nonumber
\end{eqnarray}
\noindent
and
\begin{equation}
{\rm Im}\,\Pi_{00}\boq=\frac{4\alpha\hbar c}{v_F^2\svq}
\left[\,\,\int\limits_{v_Fq-\omega}^{\infty}\!\!\!\!\dxe F_3(x)
-\!\!
\int\limits_{v_Fq+\omega}^{\infty}\!\!\!\!\dxe F_4(x)\right],
\label{eq3a}
\end{equation}
\noindent
where $\alpha=e^2/(\hbar c)$ is the fine structure constant,
$\beta=\hbar/(2k_BT)$, $k_B$ being the Boltzmann constant, and
$$F_{1,2}(x)=\sqrt{\vq-(x\pm \omega)^2},\quad
F_{3,4}(x)=\sqrt{(x\pm \omega)^2-\vq}.$$

In a similar way, for $\Pi$ one obtains \cite{54}
\begin{eqnarray}
&&
\hspace*{-5mm}
{\rm Re}\,\Pi\boq=\frac{\pi\alpha\hbar  q^2}{c}\svq+
\frac{8\alpha\hbar }{v_F^2c}\left\{
\vphantom{\left[\int\limits_{0}^{v_Fq-\omega}\dxe\right]}
\frac{\omega^2\ln2}{\beta}\right.
\label{eq4} \\
&&\hspace*{-5mm}\left. +
\frac{\svq}{2}\left[\int\limits_{0}^{v_Fq-\omega}\!\!\!\!\!\dxe
\frac{(x+\omega)^2}{F_1(x)}
-\!\!\!\!
\int\limits_{0}^{v_Fq+\omega}\!\!\!\!\!\dxe
\frac{(x-\omega)^2}{F_2(x)}\right]\right\}
\nonumber
\end{eqnarray}
\noindent
and
\begin{equation}
{\rm Im}\,\Pi\boq=\frac{4\alpha\hbar}{v_F^2c}\svq
\left[\,\,\int\limits_{v_Fq+\omega}^{\infty}\!\!\!\!\dxe
\frac{(x-\omega)^2}{F_4(x)}
-\!\!\!
\int\limits_{v_Fq-\omega}^{\infty}\!\!\!\!\dxe
\frac{(x+\omega)^2}{F_3(x)}\right].
\label{eq4a}
\end{equation}

In the remaining region of evanescent waves $v_Fq<\omega< cq$ and in
the region of propagating waves $\omega\geqslant cq$, the quantities
$\Pi_{00}$ and $\Pi$ are given by the unified expressions \cite{54}.
Thus, for $\Pi_{00}$ one has
\begin{eqnarray}
&&
\hspace*{-5mm}
{\rm Re}\,\Pi_{00}\boq=
\frac{4\alpha\hbar c}{v_F^2}\left\{
\vphantom{\left[\int\limits_{0}^{v_Fq-\omega}\dxe\right]}
\frac{2\ln2}{\beta}\right.-\frac{1}{\wsvq}
\label{eq5} \\
&&\hspace*{-5mm}\left.
\times\left[
\int\limits_{0}^{\infty}\!\!\dxe F_3(x)-\!\!\!\!
\int\limits_{\omega+v_Fq}^{\infty}\!\!\!\!\dxe F_4(x)
+\!\!\!\!\int\limits_{0}^{v_Fq-\omega}\!\!\!\!\dxe F_4(x)\right]\right\}
\nonumber
\end{eqnarray}
\noindent
and
\begin{equation}
{\rm Im}\,\Pi_{00}\boq=\frac{\alpha\hbar c}{\wsvq}
\left[\pi q^2-\frac{4}{v_F^2}\!\!\int\limits_{\,-v_Fq}^{v_Fq}\!\!\!\!dx
\frac{\sqrt{\vq-x^2}}{e^{\beta(\omega+x)}+1}
\right].
\label{eq5a}
\end{equation}

For $\Pi$ the following expressions are valid
\begin{eqnarray}
&&
{\rm Re}\,\Pi\boq=
\frac{4\alpha\hbar }{v_F^2c}\left\{
\vphantom{\left[\int\limits_{0}^{v_Fq-\omega}\dxe\right]}
\frac{2\omega^2\ln2}{\beta}\right.-\wsvq
\nonumber \\
&&
\times\left[
\int\limits_{0}^{\infty}\!\!\dxe \frac{(x+\omega)^2}{F_3(x)}-\!\!
\int\limits_{\omega+v_Fq}^{\infty}\!\!\!\!\dxe \frac{(x-\omega)^2}{F_4(x)}
\right.
\left.\left.
+\!\!\int\limits_{0}^{\omega-v_Fq}\!\!\!\!\dxe
\frac{(x-\omega)^2}{F_4(x)}\right]\right\}
\label{eq6}
\end{eqnarray}
\noindent
and
\begin{equation}
{\rm Im}\,\Pi\boq=\frac{\alpha\hbar }{v_F^2c}\wsvq
\left[-\pi\vq+4\!\!\int\limits_{-v_Fq}^{v_Fq}\!\!\!\!
\frac{dx}{e^{\beta(\omega+x)}+1}
\frac{x^2}{\sqrt{\vq-x^2}}
\right].
\label{eq6a}
\end{equation}

The polarization  tensor is gauge-invariant and, as a consequence, satisfies
the transversality condition \cite{38,39,40,41,42,43,44,45,46}
\begin{equation}
q^{\mu}\Pi_{\mu\nu}\boq=0,\,\,\, q^{\mu}=(\omega/c,q^1,q^2).
\label{eq7}
\end{equation}

Now we use an expression for the current arising due an application of the
electromagnetic field
\begin{equation}
J^{\mu}\boq=\frac{c}{4\pi\hbar}\Pi^{\mu\nu}\boq A_{\nu}\boq,
\label{eq8}
\end{equation}
\noindent
where $A_{\nu}$ is the vector potential, and the microscopic
relativistically covariant Ohm's law \cite{62a}
\begin{equation}
J^{\mu}(\omega,\mbox{\boldmath$q$})=
\sigma^{\mu\nu}(\omega,\mbox{\boldmath$q$})E_{\nu}(\omega,\mbox{\boldmath$q$}),
\label{eq12a}
\end{equation}
\noindent
{ {where $E_{\nu}$ is the 3-vector of the electric field}}
(see also Ref.~\cite{62b}
{ {for a definition of the relativistically covariant vectors of electric and
magnetic fields}}).
{ {By employing}}
Eqs. (\ref{eq8}), (\ref{eq12a}) { {and $E_{\nu}=(i\omega/c)A_{\nu}$,}}
one expresses the tensor of electric conductivity via the polarization
tensor \cite{9,60,61,62,63,64}
\begin{equation}
\sigma^{\mu\nu}\boq=\frac{c^2}{4\pi\hbar}
\frac{\Pi^{\mu\nu}\boq}{i\omega}.
\label{eq9}
\end{equation}
\noindent
With the help of this equation, the longitudinal and transverse conductivities
of graphene are presented in the form \cite{47,48,49,50}
\begin{equation}
\sigma^{\rm L}\boq=-\frac{i\omega}{4\pi\hbar q^2}\Pi_{00}\boq,
\qquad
\sigma^{\rm T}\boq=\frac{ic^2}{4\pi\hbar q^2\omega}\Pi\boq.
\label{eq10}
\end{equation}

Finally, using Eq.~(\ref{eq1}), for the longitudinal and transverse dielectric
functions of graphene we obtain \cite{47,65}
\begin{equation}
\ve^{\rm L}\boq=1+\frac{1}{2\hbar q}\Pi_{00}\boq,
\qquad
\ve^{\rm T}\boq=1-\frac{c^2}{2\hbar q\omega^2}\Pi\boq.
\label{eq11}
\end{equation}

Note that recently the basics of quantum field theoretical approach to
a description of the electric conductivity and dielectric response of
graphene were cast under doubt. It was noticed \cite{66} that some of the
results obtained using quantum field theory are in
disagreement with those following from the Kubo model.   {Based
on the Kubo formula, the polarization tensor $\Pi^{\mu\nu}$ in
Refs.~ (\ref{eq8})
and (\ref{eq9}) was replaced} with the so-called
"regularized" quantity $\widetilde{\Pi}^{\mu\nu}$ defined as  \cite{66}
\begin{equation}
\widetilde{\Pi}^{\mu\nu}\boq={\Pi}^{\mu\nu}\boq-
\lim\limits_{\omega\to 0}{\Pi}^{\mu\nu}\boq.
\label{eq12}
\end{equation}

It was shown, however, that the polarization tensor $\Pi^{\mu\nu}$ is defined
uniquely and cannot be modified with no violation of first principles
of quantum theory \cite{67}. The derivation of Eq.~(\ref{eq10}) with
$\widetilde{\Pi}_{\mu\nu}\boq$ in place of ${\Pi}_{\mu\nu}\boq$
{ {from the Kubo formula}}
in Ref.~\cite{66} used the nonrelativistic concept of causality rather than the
relativistic one as would be correct for the Dirac model.
Specifically, Ref.~\cite{66} applied the one-sided Fourier transforms from
0 to $\infty$ instead of the two-sided one from $-\infty$ to $\infty$,
which must be used in the relativistic theory, and obtained the subtracted term
in Eq.~(\ref{eq12}) making an integration by parts in the integral from 0 to $\infty$.
This resulted in  a violation of the gauge
invariance and in other physically unacceptable consequences \cite{68}.

{ {Thus, there is no any contradiction between the results obtained using the
quantum field theory and the Kubo model if the latter is applied correctly.
When using the two-sided Fourier transforms, as one should do in application
to the relativistic systems, such as graphene, the Kubo formula results in
the correct}} Eq.~(\ref{eq9})
{ {with the polarization tensor $\Pi^{\mu\nu}$}} \cite{68}.

\section{Dielectric functions of graphene at low frequencies}

Here, we consider the properties of both the longitudinal and transverse
dielectric functions of graphene in the region of strongly evanescent waves
$0<\omega<v_Fq$. Substituting Eqs.~(\ref{eq3}) and (\ref{eq3a}) in the first equality
of Eq.~(\ref{eq11}), the real and imaginary parts of the longitudinal dielectric
function are found in the form
\begin{eqnarray}
&&
{\rm Re}\,\ve^{\rm L}\boq=1+\frac{\pi\alpha c q}{2\svq}+
\frac{4\alpha c}{v_F^2q}\left\{
\vphantom{\left[\int\limits_{0}^{v_Fq-\omega}\dxe\right]}
\frac{\ln2}{\beta}\right.
\label{eq13} \\
&&\left.
+\frac{1}{2\svq}\left[\int\limits_{0}^{v_Fq-\omega}\!\!\!\!\dxe F_1(x)
-\!\!\!\int\limits_{0}^{v_Fq+\omega}\!\!\!\!\dxe F_2(x)\right]\right\}
\nonumber
\end{eqnarray}
\noindent
and
\begin{equation}
{\rm Im}\,\ve^{\rm L}\boq=\frac{2\alpha c}{v_F^2q\svq}
\left[\,\,\int\limits_{v_Fq-\omega}^{\infty}\!\!\!\!\dxe F_3(x)
-\!\!
\int\limits_{v_Fq+\omega}^{\infty}\!\!\!\!\dxe F_4(x)\right].
\label{eq13a}
\end{equation}

As is seen in Eqs.~(\ref{eq13}) and (\ref{eq13a}), in the limiting case
$\omega\to 0$ one has
\begin{eqnarray}
&&
\lim\limits_{\omega\to 0}{\rm Re}\,\ve^{\rm L}\boq=1+
\frac{\pi\alpha c}{2v_F}+
\frac{8\alpha c\ln 2}{v_F^2q}\,\frac{k_BT}{\hbar},
\nonumber\\
&&
\lim\limits_{\omega\to 0}{\rm Im}\,\ve^{\rm L}\boq=0,
\label{eq14}
\end{eqnarray}
\noindent
i.e., the longitudinal dielectric function of graphene is regular at zero
frequency. From Eq.~ (\ref{eq13a}) it is seen that ${\rm Im}\,\ve^{\rm L}\boq>0$
because the integrand in the first integral is larger than in the second
and integrated over the wider interval.

As an example, in Figure~1, ${\rm Re}\,\ve^{\rm L}$
 given by Eq.~(\ref{eq13}) is shown at $T=300$K,\hfill \\
$q=100~\mbox{cm}^{-1}${ {$\approx 1.24\times 10^{-2}~\mbox{eV}$}}
as the function of frequency in the region to the left
of the vertical dashed line
{ { $\omega =v_Fq\approx 10^{10}~$rad/s$~n\approx 6.58\times 10^{-6}~$eV}}.
When $\omega$ approaches $v_Fq$,
 ${\rm Re}\,\ve^{\rm L}$  goes to infinity.
\begin{figure}[t]
\vspace*{-10cm}
\centerline{\hspace*{-1.cm}
\includegraphics[width=6.in]{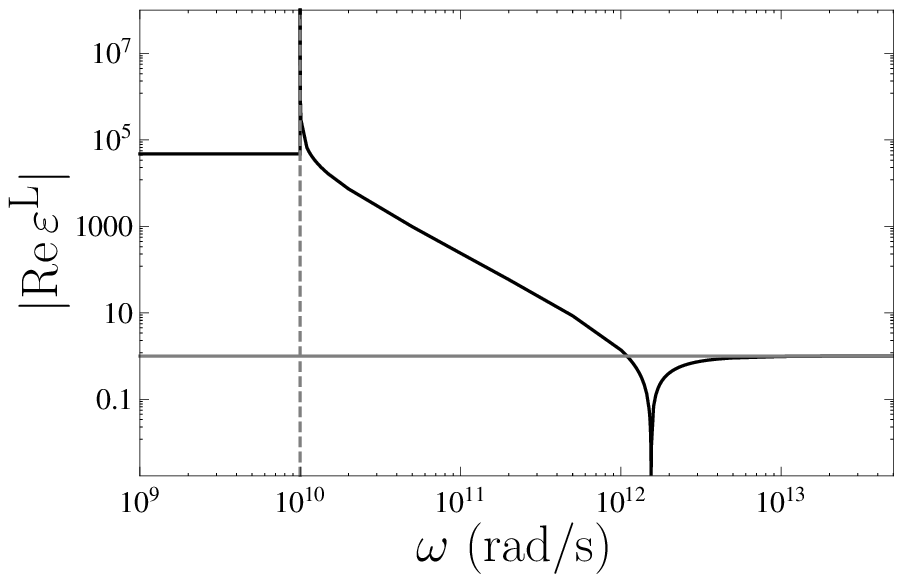}}
\vspace*{-5cm}
\caption{The magnitude of real part of the
 longitudinal dielectric function of graphene
 is shown versus frequency for $T=300$~K and
$q=100~\mbox{cm}^{-1}$ in the logarithmic scale. The threshold at $\omega=v_Fq$ is marked
by the dashed vertical line.
}
\label{figDP-1.2}
\end{figure}

 Figure.~2 shows the imaginary part of the longitudinal response function of graphene,
${\rm Im}\,\ve^{\rm L}$,
 given by Eq.~ (\ref{eq13a})  at $T=300$K,
$q=100~{\rm cm}^{-1}$ as the function of frequency in the region to the left
of the vertical dashed line $\omega=v_Fq$.
As is seen in Figure~2,  ${\rm Im}\,\ve^{\rm L}$  goes to infinity
when $\omega$ approaches $v_Fq$ from the left, as it does
 ${\rm Re}\,\ve^{\rm L}$ .
\begin{figure}[b]
\vspace*{-11cm}
\centerline{\hspace*{-1.cm}
\includegraphics[width=6.in]{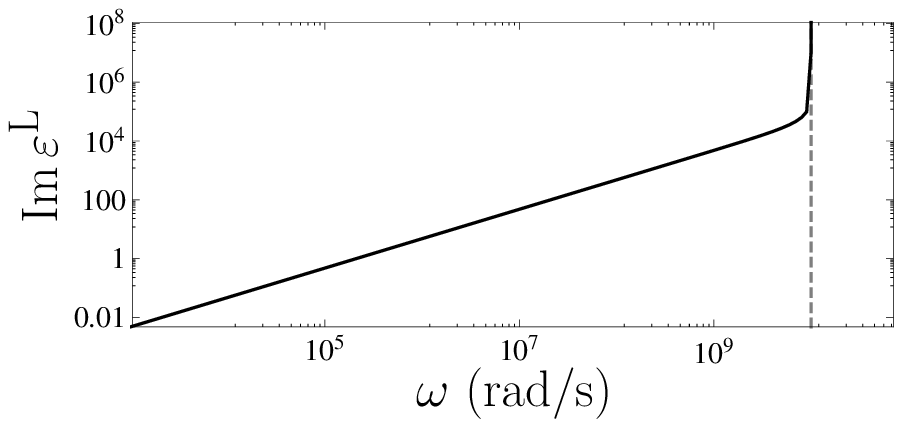}}
\vspace*{-5.cm}
\caption{The imaginary part of the
 longitudinal dielectric function of graphene
 is shown versus frequency for $T=300$~K and
$q=100~\mbox{cm}^{-1}$ in the logarithmic scale. The threshold at $\omega=v_Fq$ is marked
by the dashed vertical line.
}
\label{figDP-2.2}
\end{figure}

The real and imaginary parts of the transverse dielectric function of graphene
at low frequencies are obtained by substituting Eqs.~(\ref{eq4}) and (\ref{eq4a})
to the second equation in (\ref{eq11}). The result is
\begin{eqnarray}
&&
\hspace*{-6mm}
{\rm Re}\,\ve^{\rm T}\boq=1-\frac{\pi\alpha cq}{2\omega^2}\svq-
\frac{4\alpha c}{v_F^2q}\,\frac{\ln2}{\beta}
\label{eq15} \\
&&\hspace*{-6mm}
-\frac{2\alpha c \svq}{v_F^2q\omega^2}
\left[\int\limits_{0}^{v_Fq-\omega}\!\!\!\!\!\dxe
\frac{(x+\omega)^2}{F_1(x)}
-\!\!\!\!
\int\limits_{0}^{v_Fq+\omega}\!\!\!\!\!\dxe
\frac{(x-\omega)^2}{F_2(x)}\right]
\nonumber
\end{eqnarray}
\noindent
and
\begin{equation}
{\rm Im}\,\ve^{\rm T}\boq=\frac{2\alpha c}{v_F^2q\omega^2}\svq
\left[\,\,\int\limits_{v_Fq-\omega}^{\infty}\!\!\!\!\dxe
\frac{(x+\omega)^2}{F_3(x)}
-\!\!\!
\int\limits_{v_Fq+\omega}^{\infty}\!\!\!\!\dxe
\frac{(x-\omega)^2}{F_4(x)}\right].
\label{eq15a}
\end{equation}

Now we consider the behavior of ${\rm Re}\,\ve^{\rm T}$ and ${\rm Im}\,\ve^{\rm T}$
in the limiting case $\omega\to 0$. From (\ref{eq15}) it is seen that the second
term on the r.h.s.~behaves as $-A/\omega^2$, where $A=\pi\alpha c v_Fq^2/2$,
i.e., ${\rm Re}\,\ve^{\rm T}$ has the double pole at $\omega=0$ which is very
unusual. Recall that the commonly used   {Drude} dielectric function of
metals has the
single pole at zero frequency, whereas for dielectrics the dielectric functions
are regular at all frequencies. The formal presence of a double pole is typical
for the plasma model. { {In the case of conventional metals, it}}
is applicable only at high
frequencies belonging to the far ultraviolet and roentgen regions \cite{37}.
{ {We note that the case of a double pole
appearing in Re$\varepsilon^{\rm T}$ for graphene is not similar to the
plasma oscillations in superconductors which are described by the
dielectric permittivities possessing the double pole at zero frequency}}
\cite{68a,68b,68c,68d,68e,68f,68g,68h,68i}.
{ {The point is that the electric
current in semiconductors associated with the double pole in the dielectric
function is real and depends only on frequency in the local London limit.
By contrast, in graphene the double pole is present only at $q\neq 0$ and
the associated electric current is pure imaginary.}

The behavior of the last term in Eq.~(\ref{eq15}) in the limiting case $\omega\to 0$
is not so evident. To determine it, we introduce the parameter
$\kappa\equiv\omega/(v_Fq)$ and perform the changes of variables
$t=x/(v_Fq)\pm\kappa$ in the first and second integrals in the squared brackets,
respectively. Then one obtains
\begin{equation}
\int\limits_{0}^{v_Fq-\omega}\!\!\!\!\!\dxe
\frac{(x+\omega)^2}{F_1(x)}
-\!\!\!\!
\int\limits_{0}^{v_Fq+\omega}\!\!\!\!\!\dxe
\frac{(x-\omega)^2}{F_2(x)}=\vq(I_1+I_2),
\label{eq16}
\end{equation}
\noindent
where
\begin{eqnarray}
&&
I_1=\int\limits_{\kappa}^{1}\!\!
\frac{t^2dt}{\sqrt{1-t^2}}
\left(\frac{1}{e^{\gamma t}e^{-\gamma\kappa}+1}-
\frac{1}{e^{\gamma t}e^{\gamma\kappa}+1}\right),
\nonumber\\
&&I_2=\int\limits_{-\kappa}^{\kappa}\frac{dt}{e^{\gamma t}e^{\gamma\kappa}+1}
\,\frac{t^2}{\sqrt{1-t^2}}, \quad \gamma\equiv v_Fq\beta.
\label{eq17}
\end{eqnarray}

Under the condition $\omega\ll v_Fq$, i.e. $\kappa\ll 1$, at fixed $q,\,T\neq 0$,
we expand the integrand in $I_1$ in powers of the small parameter $\beta\kappa$
and obtain
\begin{equation}
I_1=\gamma\kappa B_1+
O(\gamma^3\kappa^3), \quad
B_1\equiv 2\int\limits_0^1 \!\!\!\frac{t^2dt}{\sqrt{1-t^2}}
\frac{e^{\gamma t}}{(e^{\gamma t}+1)^2}.
\label{eq18}
\end{equation}
\noindent
By making similar expansion in $I_2$, one finds that in the lowest order
$I_2=\kappa^3/3$ and, thus, it does not contribute to the behavior of
${\rm Re}\,\ve^{\rm T}$ at low frequencies.

As a result, substituting Eqs.~(\ref{eq16}) and (\ref{eq18}) to Eq.~(\ref{eq15}),
for the low-frequency behavior of the real part of transverse dielectric
function of graphene one finds
\begin{equation}
{\rm Re}\,\ve^{\rm T}\boq=1-8\ln 2\frac{\alpha c}{v_F^2q}\,\frac{k_BT}{\hbar}
-\frac{\alpha\hbar cv_Fq^2}{k_BT}\,\frac{B_1}{\omega}-
\frac{\pi\alpha cv_Fq^2}{2\omega^2}.
\label{eq19}
\end{equation}

Note that the "regularized" polarization tensor (\ref{eq12}) was introduced in
Ref.~\cite{66} with the aim to remove the last term in Eq.~(\ref{eq19}) which was considered
by the authors of Ref.~\cite{66} as "nonphysical". The presence of this term, however,
is in agreement with all physical principles and was confirmed experimentally
by measuring the Casimir force in graphene systems \cite{69,70}.
{ {Using the current-current correlation functions}} \cite{21}
{ { and the polarization tensor}} \cite{44},
{ {it was predicted that in the systems with a
graphene layer at nonzero temperature the Casimir force reaches the high-temperature asymptotics equal to one-half of that valid for ideal metals
already at short separations.
{{According to}} Ref.~\cite{21}, {{for graphene the high-temperature  regime
is reached under the condition $ak_BT\gtrsim \hbar v_F\approx0.0033\hbar c$.
Using the formalism of the polarization tensor,}} Ref.~\cite{44} {{indicates the
application condition of the high-temperature regime for the system of a pristine
graphene sheet parallel to a metallic plate as
$ak_BT\gg\alpha\ln\alpha^{-1}\hbar c/[2\zeta(3)]\approx 0.015\hbar c$, where
$\zeta(z)$ is the Riemann zeta function (note that}} Ref.~\cite{44}
{{uses the units with $\hbar=c=k_B=1$). Employing the more exact asymptotic
expressions for the polarization tensor, it was shown}} \cite{68j}
{{that for this system the high-temperature regime takes place under a less
severe condition
}}
\begin{equation}
ak_BT\gg\frac{\hbar v_F^2}{8\ln2~\alpha c}\approx 0.000276\hbar c.
\label{eqAs}
\end{equation}
\noindent
{{This condition is in numerical agreement with that of}} Ref.~\cite{21},
{{but, according to the condition of}} Ref.~\cite{44},
{{the high-temperature regime begins at much larger values of $ak_BT$.
The results of numerical computations}} \cite{68j} {{are in agreement
with the application condition}} \cite{21} {{of the high-temperature regime.}}
The resulting unusually big finite-temperature
Casimir effect in graphene systems
{{calculated using the polarization tensor }}
was measured in Refs.~\cite{69,70}.

Now we consider the imaginary part of $\ve^{\rm T}$ given by Eq.~(\ref{eq15a}).
By performing the same changes of variables as above,
$t=x/(v_Fq)\pm\kappa$, in the first and second integrals in the square brackets,
respectively, and expanding in powers of the small parameter $\beta\kappa$ like
this was done in the integral $I_1$ in Eq.~(\ref{eq17}), we obtain
\begin{eqnarray}
&&
\int\limits_{v_Fq-\omega}^{\infty}\!\!\!\!\dxe
\frac{(x+\omega)^2}{F_3(x)}
-\!\!\!
\int\limits_{v_Fq+\omega}^{\infty}\!\!\!\!\dxe
\frac{(x-\omega)^2}{F_4(x)}
\nonumber \\
&&
=\vq\int\limits_{1}^{\infty}\!\!
\frac{t^2dt}{\sqrt{1-t^2}}
\left(\frac{1}{e^{\gamma t}e^{-\gamma\kappa}+1}-
\frac{1}{e^{\gamma t}e^{\gamma\kappa}+1}\right)
=\vq\left[\gamma\kappa B_2+
O(\gamma^3\kappa^3)\right],
\label{eq20}
\end{eqnarray}
\noindent
where
\begin{equation}
B_2\equiv 2\int\limits_1^{\infty} \!\!\!\frac{t^2dt}{\sqrt{t^2-1}}
\frac{e^{\gamma t}}{(e^{\gamma t}+1)^2}.
\label{eq21}
\end{equation}

Substituting Eq.~(\ref{eq20}) in Eq.~(\ref{eq15a}),
the low-frequency behavior of the imaginary part of transverse dielectric
function of graphene takes the form
\begin{equation}
{\rm Im}\,\ve^{\rm T}\boq=
\frac{\alpha\hbar cv_Fq^2}{k_BT}\,\frac{B_2}{\omega}.
\label{eq22}
\end{equation}
\noindent
It is seen that the second line of Eq.~(\ref{eq20}) is evidently positive and,
thus, ${\rm Im}\,\ve^{\rm T}>0$ as it should be.

In Figure~3,  $|{\rm Re}\,\ve^{\rm T}|$  given by
Eq.~(\ref{eq15}) is shown as the function of frequency in the
region $\omega<v_Fq$ for the same values of $T$ and $q$ as in
Figures 1 and 2.
When $\omega$ approaches $v_Fq$, ${\rm Re}\,\ve^{\rm T}$ approaches
the negative constant. The
asymptotic expression (\ref{eq19}) is well applicable
at all $\omega<v_Fq$.
\begin{figure}[b]
\vspace*{-10cm}
\centerline{\hspace*{-1.cm}
\includegraphics[width=6.in]{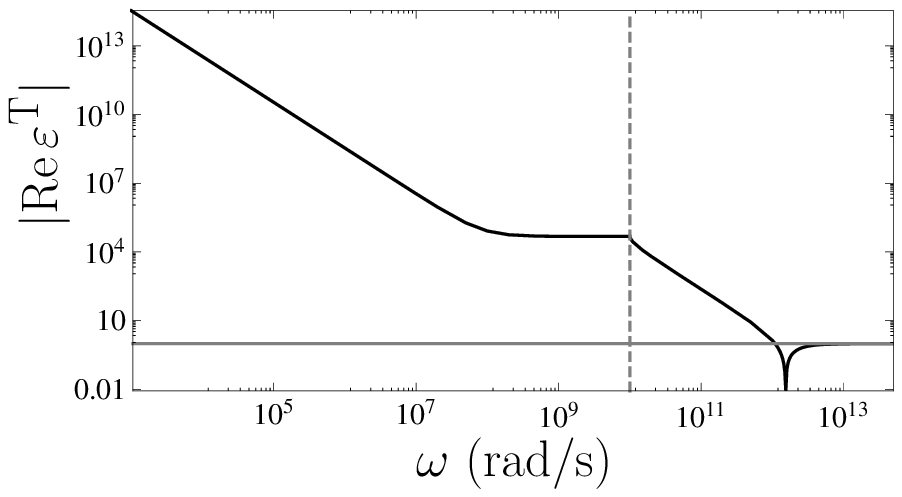}}
\vspace*{-5.cm}
\caption{The magnitude of real part of the
 transverse dielectric function of graphene
 is shown versus frequency for $T=300$~K and
$q=100~\mbox{cm}^{-1}$ in the logarithmic scale. The threshold at $\omega=v_Fq$ is marked
by the dashed vertical line.
}
\label{figDP-3.2}
\end{figure}

 Figure~4 shows the imaginary part of the transverse response function of graphene,
 ${\rm Im}\,\ve^{\rm T}$,  given by
Eq.~(\ref{eq15a}) at $T=300~$K, $q=100~\mbox{cm}^{-1}$ as the function of frequency in the
region to the left of the vertical line $\omega=v_Fq$ .
As is seen in Figure 4, ${\rm Im }\,\ve^{\rm T}$ goes to zero when $\omega$ approaches $v_Fq$
from the left.  The
asymptotic expression  (\ref{eq22}) is well applicable
for $\omega<2\times10^9$~rad/s.
\begin{figure}[t]
\vspace*{-11cm}
\centerline{\hspace*{-1.cm}
\includegraphics[width=6.in]{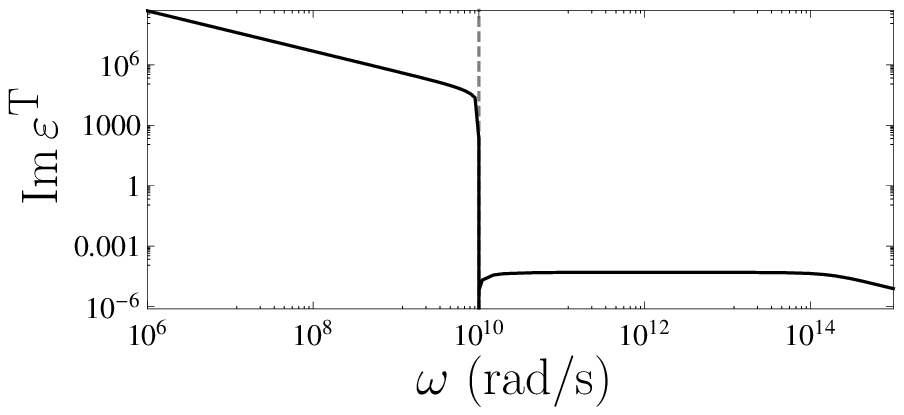}}
\vspace*{-5.cm}
\caption{The imaginary part of the
 transverse dielectric function of graphene
 is shown versus frequency for $T=300$~K and
$q=100~\mbox{cm}^{-1}$ in the logarithmic scale. The threshold at $\omega=v_Fq$ is marked
by the dashed vertical line.
}
\label{figDP-4.2}
\end{figure}

{ {As opposed to}} Eq. (\ref{eq14})
{ {for $\varepsilon^{\rm L}$, where one can consider the
limit of zero $T$,}} Eqs. (\ref{eq19}) and (\ref{eq22})
{ {for  $\varepsilon^{\rm T}$ are obtained under a condition $T\neq 0$.
The exact expressions for $\varepsilon^{\rm T}$ for any $\omega$ at
$T=0$ are}} \cite{71}
\begin{equation}
{\rm Im}\,\ve^{\rm T}\boq=\left\{
\begin{array}{cl}
-\frac{\pi\alpha qc}{\omega^2}\,\sqrt{v_F^2q^2-\omega^2}, &
\omega<v_Fq,\\[1mm]
i\frac{\pi\alpha qc}{2\omega^2}\,\sqrt{\omega^2-v_F^2q^2}, &
\omega>v_Fq.
\end{array}\right.
\label{eq29a}
\end{equation}
\noindent
{ {Thus, at zero temperature
the double pole at zero frequency in  $\varepsilon^{\rm T}$ is preserved.
Note that for graphene the limiting transitions of $\omega$ and $T$ to zero are
not interchangeable.}}
\section{Dielectric functions of graphene at high frequencies}

Now we consider the longitudinal and transverse
dielectric functions of graphene at all frequencies satisfying the condition
$\omega>v_Fq$. This includes the region of evanescent waves
$v_Fq<\omega<cq$ and the region of propagating waves $\omega\geqslant cq$.

The real and imaginary parts of the longitudinal dielectric function are
obtained by
substituting Eqs.~(\ref{eq5}) and (\ref{eq5a}) in the first equality
of  Eq.~(\ref{eq11})
\begin{eqnarray}
&&
\hspace*{-5mm}
{\rm Re}\,\ve^{\rm L}\boq=1+
\frac{2\alpha c}{v_F^2q}\left\{
\vphantom{\left[\int\limits_{0}^{v_Fq-\omega}\dxe\right]}
\frac{2\ln2}{\beta}\right.-\frac{1}{\wsvq}
\label{eq23} \\
&&\hspace*{-5mm}\left.
\times\left[
\int\limits_{0}^{\infty}\!\!\dxe F_3(x)-\!\!\!\!
\int\limits_{\omega+v_Fq}^{\infty}\!\!\!\!\dxe F_4(x)
+\!\!\!\!\int\limits_{0}^{v_Fq-\omega}\!\!\!\!\dxe F_4(x)\right]\right\}
\nonumber
\end{eqnarray}
\noindent
and
\begin{equation}
{\rm Im}\,\ve^{\rm L}\boq=\frac{\alpha c}{2q\wsvq}
\left[\pi q^2-\frac{4}{v_F^2}\!\!
\int\limits_{\,-v_Fq}^{v_Fq}\!\!\!\!dx
\frac{\sqrt{\vq-x^2}}{e^{\beta(\omega+x)}+1}
\right].
\label{eq23a}
\end{equation}

We consider first the limiting value of ${\rm Re}\,\ve^{\rm L}$
when $\omega\to\infty$. By introducing the new integration variable
$y=x-\omega$ in the second integral of Eq.~(\ref{eq23}), we obtain
\begin{equation}
-\!\!\!\!\int\limits_{\omega+v_Fq}^{\infty}\!\!\!\!\dxe F_4(x)=
-\!\!\!\int\limits_{v_Fq}^{\infty}\!\!\
\frac{dy}{e^{\beta(\omega+y)}+1}\sqrt{y^2-\vq},
\label{eq24}
\end{equation}
\noindent
which goes to zero exponentially fast when $\omega\to\infty$ and
can be omitted. In the remaining terms in the figure brackets of Eq.~(\ref{eq23}),
we change the integration variable according to $x=\omega y$, introduce the
small parameter $\delta=v_Fq/\omega$ and in the limit $\omega\to\infty$ obtain
\begin{eqnarray}
&&\hspace*{-7mm}
\frac{2\ln2}{\beta}-\frac{1}{\wsvq}
\int\limits_0^{\infty}\!\!\dxe [F_3(x)+F_4(x)]
\label{eq25}\\
&&\hspace*{-7mm}
=\frac{2\ln2}{\beta}-\frac{\omega}{\sqrt{1-\delta^2}}
\int\limits_0^{\infty}\!\!dy
\frac{\sqrt{(y+1)^2-\delta^2}+\sqrt{(y-1)^2-\delta^2}}{e^{\,\beta\omega y}+1}
=\frac{2\ln2}{\beta}-2\omega
\int\limits_0^{\infty}\!\!\frac{dy}{e^{\,\beta\omega y}+1}=0.
\nonumber
\end{eqnarray}
\noindent
Here, we took into account that when $\omega\to\infty$ it holds
$\beta\omega\gg 1$ as well. As a result, the dominant contribution
to the integrals is given by $y\ll 1$, so that
$(y+1)^2\approx( y-1)^2\approx 1$ and one can expand the square roots
in powers of the small parameter $\delta^2$.
Thus, in the limiting case $\omega\to \infty$, one obtains
\begin{equation}
\lim\limits_{\omega\to\infty}{\rm Re}\,\ve^{\rm L}\boq=1,
\quad
\lim\limits_{\omega\to\infty}{\rm Im}\,\ve^{\rm L}\boq=0.
\label{eq26}
\end{equation}
\noindent
The latter equality is an evident consequence of Eq.~(\ref{eq23a}).

{}From Eq.~(\ref{eq23a}) it follows also that ${\rm Im}\,\ve^{\rm L}>0$.
The point is that the integrand in this equation is the decreasing
function of $\beta$. Thus, it takes the maximum value for $\beta=0$
(i.e., for $T=\infty$). Then one has
\begin{equation}
\max\!\!\int\limits_{-v_Fq}^{v_Fq}\!\!\!\!dx
\frac{\sqrt{\vq-x^2}}{e^{\beta(\omega+x)}+1}=\frac{1}{2}\!
\int\limits_{-v_Fq}^{v_Fq}\!\!\!\!dx\sqrt{\vq-x^2}=
\frac{\pi}{4}\vq.
\label{eq27}
\end{equation}
\noindent
Substituting this in Eq.~(\ref{eq23a}), we find that
$\min{\rm Im}\,\ve^{\rm L}=0$ and conclude that at all $T<\infty$
it holds ${\rm Im}\,\ve^{\rm L}>0$.

In the domains of Figures 1 and 2 to the right of the dashed vertical lines ($\omega>v_Fq$),
$|{\rm Re}\,\ve^{\rm L}|$ and ${\rm Im}\,\ve^{\rm L}$, respectively, given by
Eqs.~(\ref{eq23}) and (\ref{eq23a}) are shown as the functions of frequency.
When $\omega$ increases from $v_Fq$ to $1.54\times 10^{12}$,
${\rm Re}\,\ve^{\rm L}$ varies from minus infinity to zero. With further increase of
$\omega$,  ${\rm Re}\,\ve^{\rm L}$ changes its sign and increases to unity. As to
 ${\rm Im}\,\ve^{\rm L}$, it abruptly decays to zero at $\omega>v_Fq$.

The real and imaginary parts of the transverse dielectric function of graphene
under the condition $\omega > v_Fq$ are found from Eqs.~(\ref{eq6}),
(\ref{eq6a}) and the second equality of Eq.~(\ref{eq11})
\begin{eqnarray}
&&
{\rm Re}\,\ve^{\rm T}\boq=1-
\frac{2\alpha c}{v_F^2q}\left\{
\vphantom{\left[\int\limits_{0}^{v_Fq-\omega}\dxe\right]}
\frac{2\ln2}{\beta}\right.-\frac{\wsvq}{\omega^2}
\nonumber\\
&&
\times\left[
\int\limits_{0}^{\infty}\!\!\dxe \frac{(x+\omega)^2}{F_3(x)}-\!\!
\int\limits_{\omega+v_Fq}^{\infty}\!\!\!\!\dxe \frac{(x-\omega)^2}{F_4(x)}
\right.
\left.\left.
+\!\!\int\limits_{0}^{v_Fq-\omega}\!\!\!\!\dxe
\frac{(x-\omega)^2}{F_4(x)}\right]\right\}
\label{eq28}
\end{eqnarray}
\noindent
and
\begin{equation}
{\rm Im}\,\ve^{\rm T}\boq=\frac{\alpha c }{2v_F^2q\omega^2}\wsvq
\left[\pi\vq-4\!\!\int\limits_{-v_Fq}^{v_Fq}\!\!\!\!
\frac{dx}{e^{\beta(\omega+x)}+1}
\frac{x^2}{\sqrt{\vq-x^2}}
\right].
\label{eq28a}
\end{equation}

We consider first ${\rm Re}\,\ve^{\rm T}$ in the limiting case
$\omega\to\infty$. The second integral on the r.h.s.~of Eq.~(\ref{eq28})
vanishes and the remaining two can be rearranged to
\begin{eqnarray}
&&\hspace*{-4mm}
-\frac{\wsvq}{\omega^2}\!\!\int\limits_0^{\infty}\!\!\!\dxe\left[
\frac{(x+\omega)^2}{F_3(x)}+\frac{(x-\omega)^2}{F_4(x)}\right]
\nonumber \\
&&\hspace*{-4mm}=
-\frac{\wsvq}{\omega^2}\!\!\int\limits_0^{\infty}\!\!\!\dxe\left[
F_3(x)+\frac{\vq}{F_3(x)}+F_4(x)+\frac{\vq}{F_4(x)}\right]
\nonumber\\
&&\hspace*{-4mm}
\approx -\frac{\omega}{\omega^2}\!\!\int\limits_0^{\infty}\!\!\!\dxe
(\omega+\omega)=-2\!\!\int\limits_0^{\infty}\!\!\!\dxe =
-\frac{2\ln 2}{\beta}.
\label{eq29}
\end{eqnarray}
\noindent
Thus, the first term in the figure brackets of Eq.~(\ref{eq28}) is canceled
by Eq.~(\ref{eq29}) and we obtain
\begin{equation}
\lim\limits_{\omega\to\infty}{\rm Re}\,\ve^{\rm T}\boq=1,
\quad
\lim\limits_{\omega\to\infty}{\rm Im}\,\ve^{\rm T}\boq=0
\label{eq30}
\end{equation}
\noindent
[the latter equality is evident from Eq.~ (\ref{eq28a})].

{}From Eq.~(\ref{eq28a}) it is also seen that ${\rm Im}\,\ve^{\rm T}>0$.
This is because the maximum value of the subtracted integral is
again reached at $\beta=0$ ($T=\infty$). In this case the integral  subtracted in
Eq.~(\ref{eq28a})  is
\begin{equation}
\max\!\!\int\limits_{-v_Fq}^{v_Fq}\!\!\frac{x^2dx}{[e^{\beta(\omega+x)}+1]\sqrt{\vq-x^2}}\!
=\frac{1}{2}\!\!\int\limits_{-v_Fq}^{v_Fq}\!\!\!\frac{x^2dx}{\sqrt{\vq-x^2}}\!=
\frac{\pi}{4}\vq,
\label{eq31}
\end{equation}
\noindent
which cancels the first term. Thus, at any $T<\infty$, the inequality
${\rm Im}\,\ve^{\rm T}>0$ holds.

In the region $\omega>v_Fq$  in Figures 3 and 4, $|{\rm Re}\,\ve^{\rm T}|$ and
${\rm Im}\,\ve^{\rm T}$, respectively, given by Eqs.~(\ref{eq28}) and (\ref{eq28a}) are shown
as the functions of frequency. With increasing $\omega$ from $v_Fq$ to
$1.52\times 10^{12}$~rad/s, ${\rm Re}\,\ve^{\rm T}$ varies to zero remaining
negative and then changes its sign and goes to unity. For $\omega>v_Fq$,
${\rm Im}\,\ve^{\rm T}$ increases from zero and then
decreases to zero at $\omega=\infty$. Figures 1 -- 4 demonstrate the
presence of a threshold at $\omega=v_Fq$ \cite{67,71}.

At $T=0$, below the threshold, ${\rm Re}\,\ve^{\rm {L,T}}$
are given by the first two terms in Eqs.~(\ref{eq13}) and (\ref{eq15}), whereas
${\rm Im}\,\ve^{\rm {L,T}}=0$. Above the threshold,
${\rm Re}\,\ve^{\rm {L,T}}=1$ and ${\rm Im}\,\ve^{\rm {L,T}}$
are given by the first terms in Eqs.~(\ref{eq23a}) and (\ref{eq28a}). In each
case, an order in the limiting transitions of $\omega$ and $q$ to zero
is fixed. At the point of threshold, the derivatives become discontinuous.

The dielectric functions of graphene expressed via the polarization tensor
are, by construction, analytic in the upper half-plane of complex frequency
and, thus, satisfy the Kramers-Kronig relations. The permittivity
$\varepsilon^{{\rm L}}$ is regular at zero frequency,
satisfies the standard Kramers-Kronig relations valid for dielectric
materials \cite{37}, but has a threshold at $\omega=v_Fq$. As to the permittivity
$\varepsilon^{{\rm T}}$, it is of the most nonconventional character because,
according to Eqs.~(\ref{eq19}) and
(\ref{eq22}), at nonzero temperature both the real and imaginary parts of
$\varepsilon^{{\rm T}}$ have the single pole at $\omega=0$,
whereas ${\rm Re}\varepsilon^{{\rm T}}$ also has the double
pole. As was noted above,
 both Eqs.~ (\ref{eq19}) and (\ref{eq22}) were obtained under
the condition $T \neq 0$ and it is not possible to consider the limit of zero
$T$ in these expressions. At $T=0$ there is no single pole in
$\varepsilon^{{\rm T}}$ \cite{71}.

The presence of a single pole, like that one in the imaginary part
of the dielectric permittivity of the Drude model, gives rise to the well known
additional term in the Kramers-Kronig relations \cite{37}. Similar term appears
in the case of a double pole (see Ref.~\cite{71} for details).
 In Ref.~\cite{71} it is also shown that the branch points that are present in both
$\varepsilon^{{\rm L}}$ and $\varepsilon^{{\rm T}}$
at $\omega = v_Fq$ do not affect on the form of Kramers-Kronig relations.

{ {As noted in Section 1, all derivations in this
article are made for the case of a pristine graphene possessing the zero
mass gap parameter $\Delta=2mv_F^2$, where $m$ is the quasiparticle mass,
and chemical potential $\mu$. In the case that $\Delta$ and $\mu$ are not
equal to zero, the low-frequency behavior of the dielectric functions of
graphene depends of their values preserving the same pole structure as for
a pristine graphene. Using expressions for the polarization tensor with
the arbitrary values of $\Delta$ and $\mu$ }}(see, for instance, Ref.~\cite{58}),
{ {it is easily seen that if $\Delta$, $\mu$ and $\omega$ simultaneously go to
zero one returns back to}} Eqs.~(\ref{eq14}), (\ref{eq19}) and (\ref{eq22})
{ {independently of the order of limiting transitions. }}

\section{Discussion}

In this article, we investigated the dependence of the dielectric functions of graphene
on frequency.
The most interesting unusual analytic properties were found for the real part
of the transverse function, $\varepsilon^{{\rm T}}$, at low frequencies.
Thus, both the real and
imaginary parts of $\varepsilon^{{\rm T}}$ possess the single pole at
zero frequency. What is more, the spatially nonlocal term in its real part also
possesses the double pole, which is not the case for
conventional materials according to present views.  The double pole
should be also present in the response functions of other 2D Dirac materials
such as germanene \cite{71.1,71.2,71.3}, silicene \cite{ 71.4,71.5,71.6},
phosphorene \cite{71.7,71.8,71.9}, and stanene \cite{71.10,71.11,71.12}.
In spite of   {the presence of a double pole}, the dielectric
functions of graphene satisfy all the physical demands considered above. They
possess the positive imaginary parts, which describe dissipation on the basis of
first principles, and satisfy the Kramers-Kronig relations expressing the condition
of causality. Because of this, an attempt \cite{66} to modify the polarization tensor
in order to remove the double pole predicted by the first principles of quantum
field theory is unjustified.

There is also a long-standing problem called the Casimir
puzzle.   To bring the
theoretical predictions of the fundamental
Lifshitz theory in agreement with the measurement data,
 the dielectric response of metals at low frequencies was described by
the plasma model possessing the double pole at $\omega = 0$ (see
Refs.~\cite{72,73,74,75} for a review). However, as mentioned above, this model   {is}
applicable only at high frequencies. That is why an example of graphene, whose
dielectric function possessing the double pole at zero frequency is derived starting from
first physical principles and leads to an agreement with measurements of the Casimir
force, may pave the way for resolution of the Casimir puzzle.

\section{Conclusions}

In the foregoing, we listed several  phenomenological theoretical
approaches used for investigation
of the dielectric response of graphene. It is underlined that at the characteristic
energies below approximately 3~eV the spatially
nonlocal response functions of graphene can be derived within the Dirac model
starting from first principles of thermal quantum field theory. The obtained
dielectric functions are useful for a theoretical description of many physical phenomena
in graphene systems, such as the Casimir and Casimir-Polder forces both in equilibrium situations and out of thermal equilibrium,
radiative heat transfer, atomic friction, surface plasmons etc.

According to our results, these functions possess all the properties necessary for the dielectric functions
and their transverse part has the double pole at zero frequency at any nonzero wave vector.
The above discussion { {allows to make a conjecture}}
 that the spatially nonlocal transverse electric response function of
metals possesses the double pole in the region of evanescent waves like it holds
for graphene. Recently it was demonstrated \cite{76} that the predictions of classical
electrodynamics using the Drude dielectric function for the field of oscillating
magnetic dipole reflected from a copper plate, which is fully determined by the
transverse electric evanescent waves, are in contradiction with the measurement data.
Future progress in investigation of such physical phenomena as the Casimir effect,
atomic friction, radiative heat transfer, near-field optical microscopy, total internal reflection and
 frustrated total internal reflection is closely allied to the resolution
of this problem.

This work was
supported by the State Assignment for Basic Research (project FSEG-2026-0018).

\appendix

\section{{ {The list of auxiliary functions used in the article}} }\label{A}
\setcounter{equation}{0}
\renewcommand{\theequation}{A\arabic{equation}}
{ {Here, for the readers convenience, we present a summary of the auxiliary functions
and other notations used in this article (see Table A1).
}}
\begin{table}[h]
\caption{{ {The definitions of auxiliary functions
and other notations.}}\label{tab1}}
\begin{tabular}{|c|l|}
\toprule
\hline
$\vphantom{\int\limits_0^{\infty}}$
{ {$\alpha$}}		& { {$\frac{e^2}{\hbar c}$ is the fine structure constant}} \\
\hline
$\vphantom{\int\limits_0^{\infty}}$
{ {$\beta$}}		& { {$\frac{\hbar}{2k_BT}$}} \\
\hline
$\vphantom{\int\limits_0^{\infty}}$
{ {$\gamma$}}		& { {$v_Fq\beta$}} \\
\hline
$\vphantom{\int\limits_0^{\infty}}$
{ {$\kappa$}}		& { {$\frac{\omega}{v_Fq}$}} \\
\hline
$\vphantom{\int\limits_{\frac{a}{a}}^{\frac{a}{a}}}$
{ {$B_1$}}		& { {$2\int\limits_0^1 \!\!\!\frac{t^2dt}{\sqrt{1-t^2}}
\frac{e^{\gamma t}}{(e^{\gamma t}+1)^2} $}}\\
\hline
$\vphantom{\int\limits_{\frac{a}{a}}^{\frac{a}{a}}}$
{ {$B_2$}} & { {$2\int\limits_1^{\infty} \!\!\!\frac{t^2dt}{\sqrt{t^2-1}}
\frac{e^{\gamma t}}{(e^{\gamma t}+1)^2} $}}\\
\hline
$\vphantom{\int\limits_0^{\infty}}$
{ {$F_{1,2}(x)$}}		& { {$\sqrt{v_F^2q^2-(x\pm\omega)^2}$}} \\
\hline
$\vphantom{\int\limits_0^{\infty}}$
{ {$F_{3,4}(x)$}}		& { {$\sqrt{(x\pm\omega)^2-v_F^2q^2}$}} \\
\hline
$\vphantom{\int\limits_{\frac{a}{a}}^{\frac{a}{a}}}$
{ {$I_1$}} &	{ {$\int\limits_{\kappa}^{1}\!\!
\frac{t^2dt}{\sqrt{1-t^2}}
\left(\frac{1}{e^{\gamma t}e^{-\gamma\kappa}+1}-
\frac{1}{e^{\gamma t}e^{\gamma\kappa}+1}\right)$}}	   \\
\hline
$\vphantom{\int\limits_{\frac{a}{a}}^{\frac{a}{a}}}$
{ {$I_2$}}&{ {$\int\limits_{-\kappa}^{\kappa}\frac{dt}{e^{\gamma t}e^{\gamma\kappa}+1}
\,\frac{t^2}{\sqrt{1-t^2}}$}}\\
\toprule
\end{tabular}
\end{table}


\end{document}